# Integer and Fractional Quantum Hall Effect in Two-Terminal Measurements on Suspended Graphene.

I. Skachko[1], X. Du[1]*, F. Duerr[1], A. Luican[1], D. A. Abanin[2],

L. S. Levitov[3] and E.Y.Andrei[1]

[1]Department of Physics and Astronomy, Rutgers University, Piscataway, NJ 08855

[2]Princeton Center for Theoretical Science, Princeton University, Princeton, NJ 08544

[3]Department of Physics, Massachusetts Institute of Technology, 77 Massachusetts Ave, Cambridge, MA 02139

Abstract

We report the observation of the quantized Hall effect in suspended graphene probed with a two-terminal lead geometry. The failure of earlier Hall-bar measurements is discussed and attributed to the placement of voltage probes in mesoscopic samples. New quantized states are found at integer Landau level fillings outside the sequence $\pm 2, \pm 6, \pm 10..$, as well as at a fractional filling ν=1/3. Their presence is revealed by plateaus in the two-terminal conductance which appear in magnetic fields as low as 2 Tesla at low temperatures and persist up to 20 Kelvin in 12 Tesla. The excitation gaps, extracted from the data with the help of a theoretical model, are found to be significantly larger than in GaAs based electron systems.

PACS numbers: 73.43.-f, 73.90f

* Present address: Dept of Physics, Stony Brook, NY

Since its first isolation [1] graphene, a one-atom thick membrane of crystalline carbon, has attracted growing interest[2] due to its unusual electron properties, such as a record-high mobility[3] and the anomalous integer quantum Hall effect (IQHE) [4] which was shown to persist up to room temperature[5]. Strong Coulomb interactions between electrons expected in this material have led researchers to predict new types of collective behavior[6-12], but until now experimental evidence of such behavior has been conspicuously missing. In this Letter we demonstrate that when graphene is suspended and thus isolated from substrate-induced perturbations[13] it does in fact host many-body electron states, revealed in our experiment by the observation of integer and fractional quantization of the two-terminal conductance. The IQHE is a property of non-interacting 2d electron-systems (2DES) in a perpendicular magnetic field, $B$, marked by plateaus in the transverse (Hall) conductance, $\sigma_{xy} = \nu e^2/h$, where $e$ is the electron charge, $h$ is Planck's constant, and $\nu = \frac{n_s h}{eB}$ is the filling factor. The carrier density, $n_s$, is tuned by the gate voltage; in our samples[13] $n_s[\text{m}^{-2}] \sim 2\times10^{14}V_g[\text{V}]$. In ideally clean graphene the neutrality point, $V_g = 0$, corresponds to the Dirac point, $n_s = 0$, and the carrier charge changes sign from positive (holes) to negative (electrons) as $V_g$ is swept through zero. In the absence of interactions the IQHE plateaus occur at $\nu = \pm 4(n+1/2); \quad n = 0,1,2...$, where the prefactor accounts for the 4-fold spin and valley degeneracy of Landau levels, the plus/minus sign reflects the electron-hole symmetry, and the ½ shift is a unique property of the relativistic charge carriers in graphene [2,4]. Lifting of the Landau level degeneracy due to Coulomb interaction or Zeeman coupling can lead to additional IQHE states outside this sequence. In addition, strong correlations between the electrons in partially filled Landau levels can give rise to the fractional quantum Hall effect (FQHE) seen as plateaus at fractional filling factors[14].

Thus far the IQHE in graphene was only observed in samples prepared by mechanical exfoliation of graphite onto a Si substrates capped with $SiO_2$. In these samples the non-interacting IQHE is quite robust, persisting to fairly low fields and high temperatures[5]. In contrast, the IQHE plateaus outside the non-interacting sequence could only be observed in very high fields, exceeding 25T for $\nu = \pm 1, \pm 4$, while the FQHE has not been observed in such samples. Because the FQHE state can be suppressed even by minute amounts of disorder its observation in the GaAs-based 2DES became possible only after achieving high material quality [14]. This raises an important question: does its absence in graphene reflect a fundamental limitation on correlated states of relativistic charge-carriers or is it a consequence of imperfect sample quality.

A significant improvement in sample quality was recently achieved in suspended graphene (SG) devices, where substrate-induced perturbations are eliminated [13,15]. In these samples the carrier motion approaches ballistic transport on micron-size length scales. In addition, due to the more uniform spatial distribution of the carrier density, the physics at the Dirac point can be probed with better accuracy than in non-suspended samples. The combination of ballistic transport and low carrier density achieved in SG has raised hopes to observe correlated behavior of relativistic electrons in the quantized Hall regime. However, magneto-transport measurements in SG using the standard Hall bar lead geometry have been disappointing. Not only was the FQHE not observed but even the normally much more robust IQHE could not be seen [15]. In contrast, early measurements of SG samples using a 2-terminal lead geometry did exhibit a plateau at $\nu = 2$ corresponding to the IQHE [13].

In the present study the magneto-transport of SG devices was investigated in both two-terminal and Hall bar lead geometries. The SG devices were fabricated from conventional devices, mechanically exfoliated onto Si/SiO$_2$ substrates, by removing the SiO$_2$ layer with chemical etching. In the final device the graphene sample is suspended from the Au/Ti leads (fig. 1a and inset Fig. 2a). Following fabrication the samples were baked at 200 $^0$C in flowing forming gas (Ar:H$_2$ 9:1) for two hours and current annealed[16] in vacuum at low temperature to further improve sample quality. To ensure mechanical integrity we focused on short devices, <1μm long. Longer devices tend to collapse during fabrication or when subject to a gate voltage.

In Fig. 1 we show the results obtained for an SG device with a Hall bar lead geometry. The filling factor dependence of the Hall ($G_{xy}$) and longitudinal ($G_{xx}$) conductance is shown in Fig. 1b. Surprisingly, $G_{xy}$ does not show clear plateaus and its value is significantly lower than that expected for the QHE plateaus, $\nu e^2/h$. At the same time $G_{xx}$ instead of vanishing for ν corresponding to the QHE, develops plateau-like features. This indicates that, although the QH effect does occur in our system, the Hall voltage probes are short-circuited through the source and drain leads. To understand these data we consider the potential distribution obtained from the magnetotransport problem solved for parameters comparable to the experiment (Fig. 1c). At large Hall angles (the case of quantum Hall plateaus) most of the electric field is concentrated at two accumulation regions at opposite corners of the sample also known as "hot spots" in the Hall effect literature [17-19]. Since current transport in this regime (in the simplest picture) occurs in ballistic chiral edge channels (Fig. 1d), carriers in the top (bottom) edge channel originating from the source (drain) necessarily traverse the hot spot region as they enter the opposite current lead. A voltage probe placed within this region senses a potential whose value is intermediate between that of the edge channel and the current lead. This can cause a significant reduction in the measured Hall voltage between opposite voltage probes. This effect can be quite significant in mesoscopic samples where the lead dimensions are comparable to sample size. Thus, it is likely that the shorting of the Hall voltage observed in SG devices is due to samples that are too small to place voltage leads outside the hot spot regions.

Although at present reliable Hall-bar measurements are not feasible in SG, it is still possible, as we show below, to obtain the Hall conductance on QH plateaus as well as the longitudinal conductivity, $\sigma_{xx}$, from a two-terminal measurement, where the leads completely avoid the hot spots. The two-terminal conductance $G$ measured in our SG devices in the ballistic regime is shown in Fig. 2. From the carrier density dependence of the zero field resistivity (Fig. 2a) we extract the Drude mobility, $\mu = (\rho n e)^{-1}$ and find that its density dependence, $\mu \propto n_s^{-1/2}$, closely follows that of a ballistic device with the same dimensions (Fig. 2a lower inset). For $n_S = 10^{10}$cm$^{-2}$ the mobility reaches 260,000 cm$^2$/Vs and increases further as the density is reduced. The mean free path, obtained from the Einstein relation, is about 250 nm. This is nearly half the distance between the leads in our device, indicating a regime of nearly ballistic transport[13]. Accordingly, the zero-field density dependence of $G$ (Fig. 2a) is close to the theoretical prediction for ballistic transport [20] on the hole side of the Dirac point, with a somewhat less good agreement on the electron side. In finite field the conductance exhibits well-defined plateaus, $G = |\nu|e^2/h$ for $\nu = \pm 2, \pm 6, \pm 10, \pm 14...$ seen already below 1 Tesla (Fig. 2b). In between the plateaus we observe conductance maxima in agreement with the expectation for our device geometry ($W/L>1$) [21,22]. When $G$ is plotted versus the filling-factor the curves for all fields lie on top of each other (Fig 2c) and exhibit a series of well defined IQHE plateaus, indicating high carrier mobility [21] and negligible contact resisatnce.

Above 2 Tesla additional plateaus develop at $\nu=\pm 1, 3$ reflecting interaction induced lifting of the degeneracy in the $n = 0$ and $n = 1$ Landau levels. This is in sharp contrast to earlier measurements on non-suspended samples where the $\nu = \pm 1$ plateaus were only seen above 26T [23] and the one at $\nu=3$ was absent. Remarkably, at low temperature we observe a FQHE plateau at $\nu= -1/3$ with $G=(1/3)e^2/h$ which becomes better defined with increasing field (Fig 2b). Zooming into the low $\nu$ regime in Fig. 3a we note that the plateaus at $\nu= -1/3, -1$ show accurate values of the quantum Hall conductance. Crucially, the densities at which these plateaus occur scale linearly with the $B$ field, $n_s=\nu eB/h$, corresponding to constant filling factor values. As seen in Fig. 3b, the FQHE at $\nu = -1/3$ is quite robust; it appears already at ~ 2 Tesla at low temperatures and persists up to ~ 20K at 12 Tesla. At even lower filling factors the conductance becomes vanishingly small indicating that the FQHE in graphene competes with an insulating phase at $\nu = 0$. The properties of this phase including temperature dependence and activation gaps are discussed in reference [24.

In non relativistic 2DES, the FQHE at $\nu=1/3$ reflects the formation of an incompressible Laughlin-type condensate. Elementary excitations in this FQHE state are quasiparticles with fractional charge $e/3$ and a finite energy gap $\Delta>0$. The gap can be estimated as $\Delta=\alpha e^2/4\pi\varepsilon_0\varepsilon l_c$, with $\varepsilon_0$ the permittivity of free space, $\varepsilon$ the effective dielectric constant, $l_c = (\hbar/eB)^{1/2}$ the magnetic length, and $\alpha$ a numerical constant, typically of order 0.1 according to theory but much smaller (~0.01) in actual measurements[25]. It is not obvious that the FQHE in graphene represent the same physical state. In fact the FQHE in graphene is expected to deviate in important ways from that in GaAs-based 2DES. First, electrons in graphene are considerably more two-dimensional than in GaAs quantum wells, where the well widths range from 10 to 30 nm. This makes the interaction at short distances in graphene much stronger than in quantum wells. The interactions in SG are further enhanced due to the absence of substrate screening ($\varepsilon \sim 1$) in contrast to GaAs where $\varepsilon \sim 13$. Stronger interaction leads to a larger energy gap [6], and thus higher temperatures at which the FQHE can be observed. Another difference arises from the four-fold spin and valley degeneracy. Because of this degeneracy, the situation in graphene is more similar to that realized in double-quantum-wells rather than in single-quantum-well systems. However, electron interactions in graphene correspond to nearly identical intra- and inter-well interactions, a regime nearly impossible to achieve in GaAs systems. This leads to an *SU(4)* symmetry of the interaction Hamiltonian, and the prediction of new FQHE states with no analog in GaAs [3,8].

In order to gain insight into the FQHE state in graphene, it is desirable to obtain an accurate measure of the excitation gap. Experimentally the gap is usually extracted from the temperature dependence of $\sigma_{xx}$ which can be obtained, together with $\sigma_{xy}$, from a multi-probe measurement such as a Hall bar lead geometry. However, as discussed above, this method fails in mesoscopic samples.

Since at present there are no reliable Hall-bar measurements in SG, here we attempt to obtain $\sigma_{xx}$ and $\sigma_{xy}$ from the two-terminal conductance *G*. Since this quantity depends on $\sigma_{xx}$, $\sigma_{xy}$ and sample geometry, 'deconvolving' *G* requires an additional input. Such input is provided by the conformal invariance of the magnetotransport problem [21]. In this approach, $\sigma_{xx}$ and $\sigma_{xy}$ are interpreted as a real and imaginary part of a complex number $\sigma = \sigma_{xx} + i\sigma_{xy}$, and thereupon the transport equations become conformally invariant. Applied to a rectangular two-lead geometry, such as that in our experiment, theory [21] yields a specific dependence of *G* on $\sigma_{xx}$, $\sigma_{xy}$ and the aspect ratio *W/L*. Interestingly, the same dependence describes two-terminal conductance for an arbitrary sample shape, with the "effective aspect ratio" *W/L* encoding the geometry dependence.

There are several ways to use this approach for determining $\sigma_{xx}$. One is to focus on the plateaus, where $\sigma_{xx}$ is small and $\sigma_{xy}$ is quantized, $\sigma_{xy} = ve^2/h$. Expanding *G* in the small ratio $\sigma_{xx}/|\sigma_{xy}| \ll 1$, the deviation from a quantized value can be expressed as $G = |v|e^2/h + k \sigma_{xx}$. Theory [21] yields the coefficient *k* which, as a function of *W/L*, is positive

(negative) for *L<W* (*L>W*), and zero for *L=W*. Despite its conceptual simplicity, we found this approach difficult to implement, since the effective *W/L* may significantly deviate from the geometric aspect ratio of the sample, and thus should be treated as a fitting parameter [22].

Considerably more reliable results can be obtained by focusing on the N-shaped distortions of the plateaus [21, 22]. The N-shaped features can be described by the density-dependent $\sigma_{xx}(\nu)$ and $\sigma_{xy}(\nu)$ obtained from the so-called semicircle relation (see [26] and references therein). This relation, which is also rooted in the complex-variable interpretation of magnetotransport, is known to work well in GaAs-based 2DES both in the IQHE and the FQHE regimes [27,28]. For a plateau-to-plateau transition between incompressible filling factors $\nu_1 < \nu_2$ the semicircle relation gives $\sigma_{xx}^2 = (\sigma_{xy} - \nu_1)(\nu_2 - \sigma_{xy})$ (in units of $e^2/h$). For fitting the conductance data shown in Fig.3, with the incompressible states at $\nu=0, -1/3, -1, -2$, we model the contribution of each plateau-to-plateau transition by a Gaussian $\sigma_{xx}(\nu) = \frac{1}{2}(\nu_2 - \nu_1)e^{-A(\nu-\nu_c)^2}$, $\nu_1 < \nu_c < \nu_2$, and find the corresponding $\sigma_{xy}$ from the semicircle relation. This gives a contribution to the longitudinal and Hall conductivity of each of the relevant Landau (sub) levels. The net conductivity, found as a sum of such independent contributions, is then used to calculate *G* (ν) using the approach [21] as discussed above.

We treat the $\sigma_{xx}$ peak positions and widths, as well as the effective ratio *W/L*, as variational parameters. As illustrated in Fig.4a, the Gaussian model with individually varying peak widths and positions is found to provide a rather good description of the data, at the same time giving the quantity $\sigma_{xx}(\nu)$. The approach based on treating *W/L* as a variational parameter, in general different from the actual sample aspect ratio, was shown to work rather well in the IQHE regime [22]. It was conjectured [22] that variations in the best-fit value of *W/L* account for the sample-dependent specifics of the current flow pattern such as those due to imperfect contacts and/or contact doping.

In order to extract $\sigma_{xx}$ at the plateaus $\nu = -1/3$ and $\nu = -1$, we analyzed a set of fits which best follow the data near these filling factors (Fig. 4a). In both cases we found an optimal effective aspect ratio $W/L \approx 1.7$ (such deviation from the geometric aspect ratio, which for our sample is close to 2.5, is consistent with the results of Ref.[22]). Best fits were found from the standard deviation in the conductance averaged over the range of densities on the plateau and around it; in the case of the 1/3 plateau we excluded a small interval near $\nu \approx 0.4$ which we believe is related to another incipient FQHE feature.

The value of $\sigma_{xx}$ at the minima of $\sigma_{xx}(\nu)$ was taken as an estimate of the longitudinal conductivity of the incompressible QH states. Statistical error was estimated from the spread in the $\sigma_{xx}$ values found from the fits with standard deviation within 30% of the best fit. The resulting error bar, displayed in Fig. 4b, is below 10% at 10K, and increases to 20-25% at 1.2K as a result of pronounced mesoscopic fluctuations developing at low temperatures.

We used the 12 Tesla data at $T$=1.2, 4.2, 6 and 10 K to infer the temperature dependence of $\sigma_{xx}$ in the $\nu = -1/3$ state. The results, displayed in Fig.4b, were analyzed in terms of scaling with the function $\log \sigma_{xx}^{-1} \approx (T^*/T)^\eta$. The data were modeled with both the activation dependence, $\eta = 1$, and the one-dimensional variable-range-hopping dependence, $\eta = 1/2$. For the 1/3 state we find that $\eta = 1/2$, with characteristic temperature $T^* \sim 7$K, gives the best fit corresponding to ~3% of the theoretically predicted value for $\varepsilon$=1[6-8]. Similar hopping conduction behavior was found to describe the low temperature FQHE transport in semiconductor samples[25]. For the $\nu = -1$ state we find that activated behavior, $\sigma_{xx} \sim \exp(-\Delta/2k_BT)$, with the energy gap value $\Delta/k_B$= 9.6K, best fits the data. This corresponds to ~ 7% of the predicted value for $\varepsilon$=1 [6-8].

In summary, this work demonstrates that when graphene is isolated from environmental disturbances caused by the substrate, the collective behavior of the relativistic 2D carriers become observable. By using two-terminal magneto-transport measurements we showed that SG samples exhibit a robust FQHE at $|\nu|$=1/3 and found that the energy scale of the quasiparticle excitations is larger than that for 2DES in semiconductors, reflecting the stronger Coulomb interactions in graphene.

EYA acknowledges DOE support under DE-FG02-99ER45742 and partial support under NSF-DMR-045673. LSL acknowledges ONR support under N00014-09-1-0724. We thank J. Jain, A. Akhmerov, S. DasSarma, V. Falko, M. Fuhrer and A. Geim for useful discussions.

Captions:

Figure 1. **a.** SEM image of SG sample with Hall bar lead geometry. **b.** Longitudinal and Hall conductance for the sample in (a) obtained from the measured resistances $G_{xy} = R_{xy}/(R_{xy}^2 + R_{xx}^2), G_{xx} = R_{xx}/(R_{xy}^2 + R_{xx}^2)$ . **c.** Equipotential line distribution in a sample with $W/L=1.78$, for a large Hall angle ($\sigma_{xy}/\sigma_{xx} = 20$), illustrating the regions of sharp potential drop (hot spots) at opposite corners of the sample. **d**. Shorting of Hall voltage by invasive leads. In the quantized Hall regime the carriers move in ballistic edge channels originating from one current lead and traversing a region of potential gradient, "hot spot" (purple circle), before entering the opposite lead. A voltage lead covering the hot spot region (hashed rectangles) measures a reduced Hall voltage as observed in (b).

Figure 2. **a.** Comparison of measured density dependence of resistivity in zero field (red) with the calculated curve (black) for an ideal ballistic graphene junction with the same dimensions [20]. Upper inset: false-color SEM image of SG device with $L= 0.6$ μm and $W = 1.4$μm. Lower inset: comparison of measured (red) and calculated 9black) density dependence of mobility. **b.** Gate voltage dependence of two-terminal resistance at $T=1.2$K for the indicated field values. Dashed lines mark the resistance values on the quantized Hall plateaus. The IQHE plateau at $\nu= -1$ and the FQHE plateau at $\nu= -1/3$ are clearly visible on the hole side ($V_g < 0$). **c.** Conductance $G$ vs. scaled electron density, $\nu=n_s h/eB$, in (b). All the curves lie on top of each other, exhibiting plateaus at $G(\nu)=|\nu|e^2/h$ centered at the correct filling factors in the hole sector. The N-shape distortions characteristic of the two-terminal measurements with $L<W$ are clearly seen[21]. **d.** Evolution of the plateaus in resistance with magnetic field for indicated values of gate voltage for an SG sample with $L=2.6$ μm, $W=0.5$μm. Horizontal lines indicate theoretical resistance values for quantized Hall plateaus.

Figure 3. Field and temperature dependence of the $\nu= -1/3, -1$ plateaus. **a.** Zoom into the low $\nu$ segment of the data in Fig. 2c. Well defined conductance plateaus at $G = e^2/h$ and $G = 1/3(e^2/h)$ occurring at the precise filling factors $\nu=-1, -1/3$ are already visible at

2T. **b.** Temperature dependence of the quantized Hall plateaus. On the plateaus, the conductance increases with temperature faster than off the plateaus. Both plateau features remain visible up to $T\sim 20$K.

Figure 4. **a.** Theoretical fit of the two-terminal conductance (red) shown together with the data (black). The best fit was obtained using the semicircle model and treating the effective aspect ratio $W/L$ as a fitting parameter as described in the text. The longitudinal conductivity $\sigma_{xx}$ (blue), obtained from the best fit, has peaks at the plateau-to-plateau transitions and minima on the plateaus. Values at the minima are used to estimate $\sigma_{xx}$ for incompressible IQHE and FQHE states. **b.** Fits of the temperature dependence of $\sigma_{xx}$ at $\nu = -1/3, -1$ to models of activation and variable-range-hopping (VRH). For $\nu = -1/3$ the VRH model, $\sigma_{xx} \sim \exp(-(T^*/T)^{1/2})$, gives the best fit, with energy scale $T^* \sim 7$ K. For $\nu = -1$ the activation model, $\sigma_{xx} \sim \exp(-\Delta/2k_B T)$, gives the best fit with $\Delta/k_B \sim 9.6$ K.

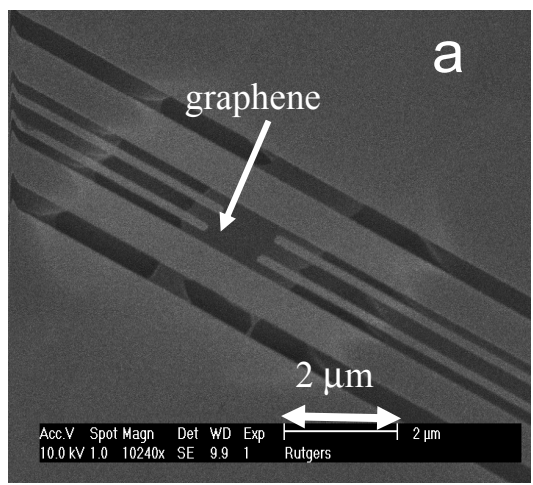
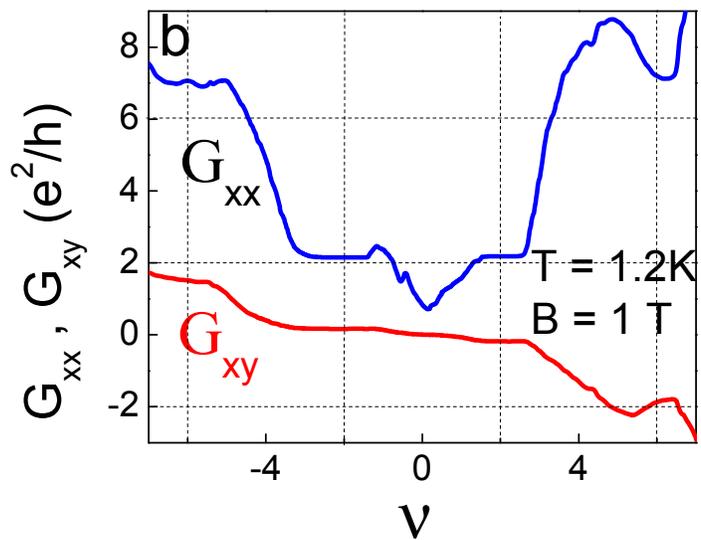
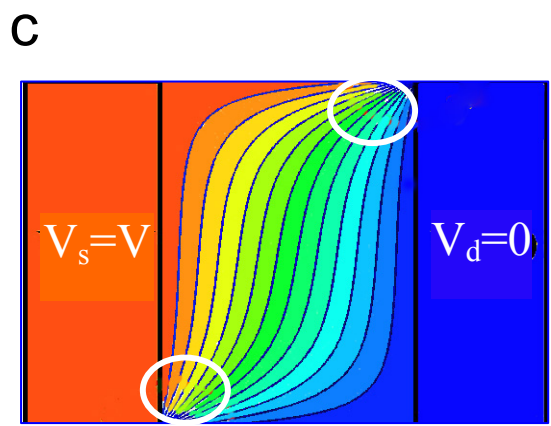
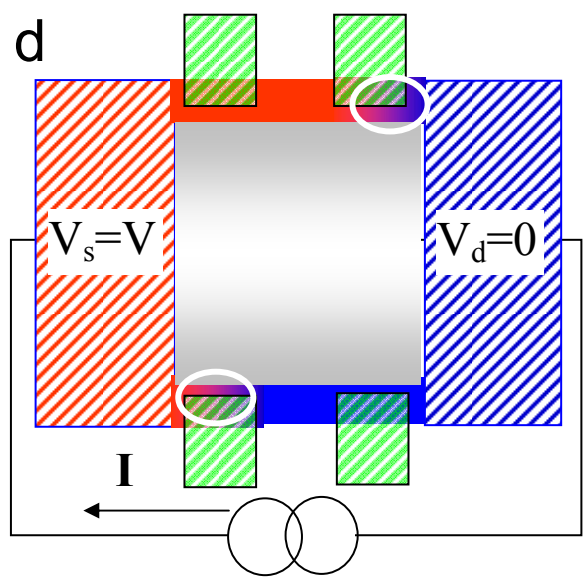

Figure 1

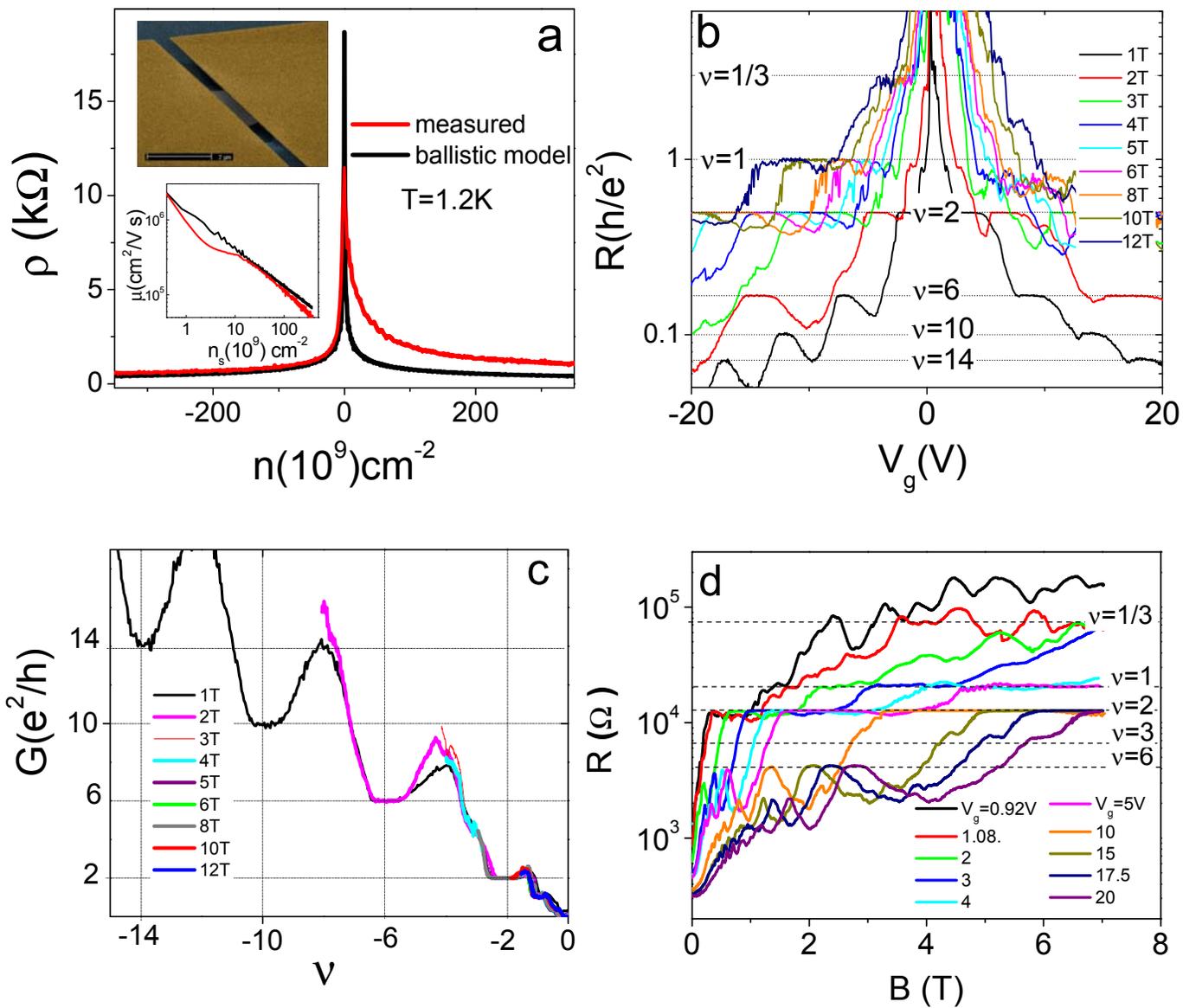

Figure 2

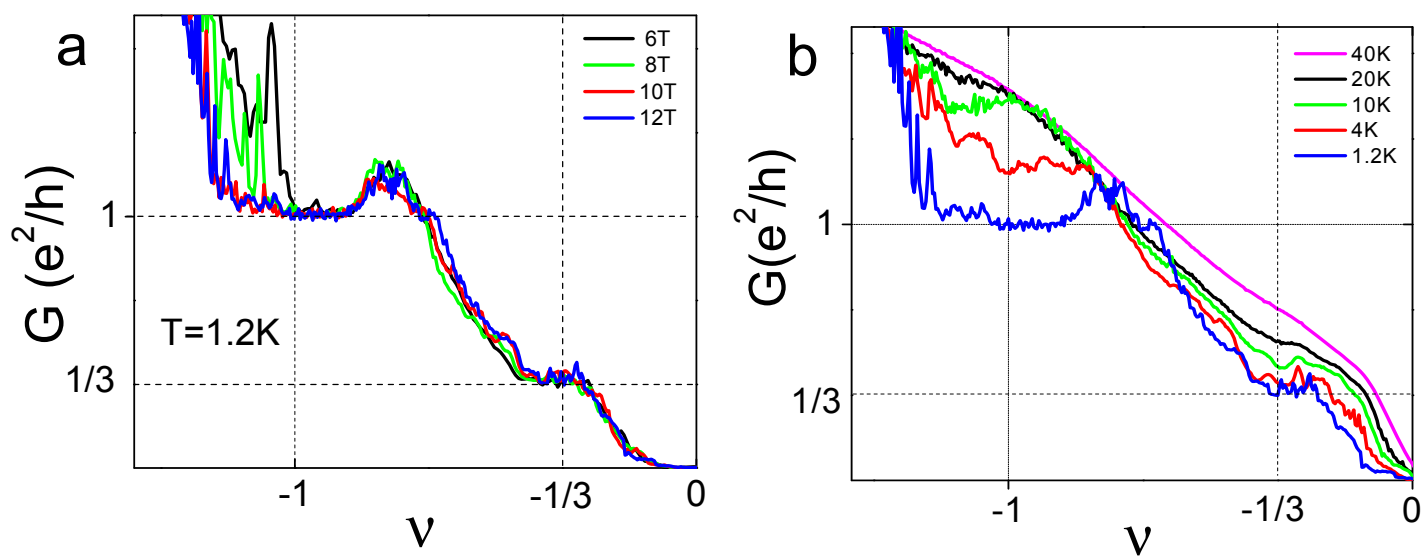

Figure 3

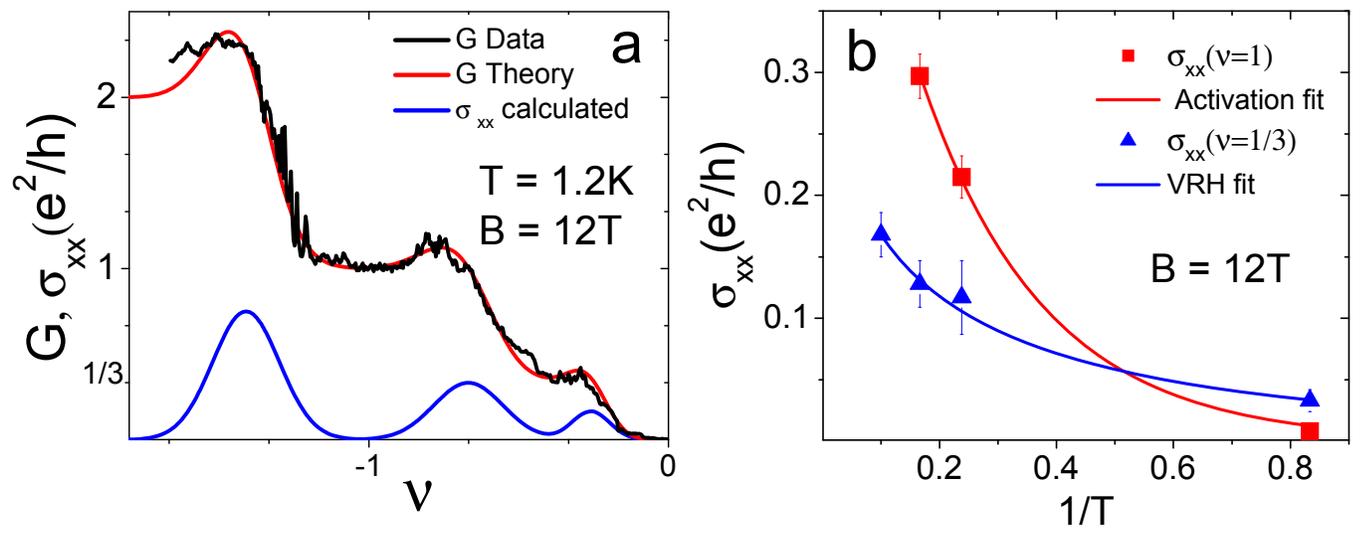

Figure 4